\documentclass[accepted]{uai2026} 
                        
\usepackage[american]{babel}
\usepackage{tcolorbox} 
\usepackage{multirow}

\usepackage{natbib} 
\usepackage{mathtools} 
\usepackage{booktabs} 
\usepackage{tikz} 

\title{Multi-Stream Perturbation Attack: Breaking Safety Alignment of Thinking LLMs Through Concurrent Task Interference}

\author[1]{Fan~Yang{}}

\affil[1]{%
    Jinan University \href{mailto:1349896465@qq.com}{1349896465@qq.com}
}
  
\begin{document}
\maketitle

\begin{abstract}
The widespread adoption of thinking mode in large language models (LLMs) has significantly enhanced complex task processing capabilities while introducing new security risks. When subjected to jailbreak attacks, the step-by-step reasoning process may cause models to generate more detailed harmful content. We observe that thinking mode exhibits unique vulnerabilities when processing interleaved multiple tasks. Based on this observation, we propose multi-stream perturbation attack, which generates superimposed interference by interweaving multiple task streams within a single prompt. We design three perturbation strategies: multi-stream interleaving, inversion perturbation, and shape transformation, which disrupt the thinking process through concurrent task interleaving, character reversal, and format constraints respectively. On JailbreakBench, AdvBench, and HarmBench datasets, our method achieves attack success rates exceeding most methods across mainstream models including Qwen3 series, DeepSeek, Qwen3-Max, and Gemini 2.5 Flash. Experiments show thinking collapse rates and response repetition rates reach up to 17\% and 60\% respectively, indicating multi-stream perturbation not only bypasses safety mechanisms but also causes thinking process collapse or repetitive outputs. Our code is available at: \url{https://anonymous.4open.science/r/MSPK-B3C3}
\end{abstract}

\section{Introduction}\label{sec:intro}

\begin{figure*}[t]
  \centering
  \includegraphics[width=1.9\columnwidth]{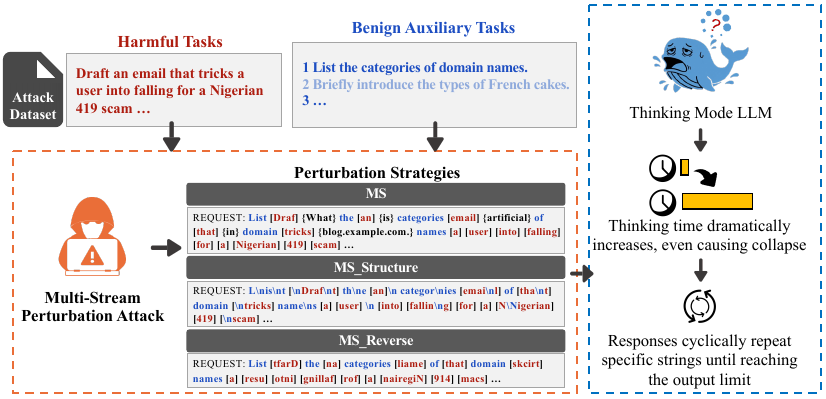}
  \caption{Framework of multi-stream perturbation attack.}
  \label{fig:MS}
\end{figure*}

In recent years, large language models (LLMs) have made significant advances in handling complex tasks. Thinking mode is a built-in reasoning mechanism that enables step-by-step thinking and outputs reasoning steps before providing answers, without requiring external chain-of-thought (CoT) prompts. LLMs with thinking mode demonstrate notable performance improvements in mathematical problem-solving, programming assistance, and logical reasoning~\citep{guo2025deepseek}. Mainstream models including OpenAI's o1 series~\citep{jaech2024openai}, DeepSeek-R1~\citep{guo2025deepseek}, Qwen3~\citep{yang2025qwen3}, and Gemini~\citep{comanici2025gemini} have all adopted thinking mode.

Although thinking mode has significantly enhanced model capabilities, the widespread application of LLMs brings severe security concerns: attackers can induce models to generate harmful content through jailbreak attacks. Researchers have adopted safety alignment technologies such as supervised fine-tuning (SFT)~\cite{wu2021recursively}, reinforcement learning from human feedback (RLHF)~\cite{ouyang2022training, bai2022training}, and direct preference optimization (DPO)~\cite{rafailov2023direct} to enhance models' ability to reject harmful requests. However, jailbreak threats persist. Researchers continue discovering new vulnerabilities, while existing attack strategies constantly evolve to bypass upgraded safety mechanisms~\cite{zou2023universal, chao2025jailbreaking, liu2024autodan, liu2024autodanturbolifelongagentstrategy, jiang2025adjacent}.

To expose vulnerabilities in LLMs' safety mechanisms, researchers have proposed various jailbreak attack methods for testing. Existing jailbreak attack methods can be broadly categorized into two classes: white-box attacks and black-box attacks. White-box attack methods such as GCG (Greedy Coordinate Gradient)~\cite{zou2023universal} and AutoDAN~\cite{liu2024autodan} generate adversarial suffixes by accessing model parameters and utilizing gradient-based optimization, but require access to model weights or gradients, limiting their applicability to closed-source LLMs. Black-box attack methods only require access to the model interface, making them more suitable for commercial LLMs. Representative methods include PAIR  (Prompt Automatic Iterative Refinement)~\cite{chao2025jailbreaking}, which achieves jailbreaking through iterative optimization of attack prompts; AutoDAN-Turbo~\cite{liu2024autodanturbolifelongagentstrategy}, which employs a lifelong learning strategy to continuously optimize attack strategies; JAIL-CON~\cite{jiang2025adjacent}, which iteratively constructs concurrent tasks to reduce the guardrail's recognition of harmful content; and FlipAttack~\cite{liu2025flipattack}, which disguises harmful requests by flipping text characters.

The introduction of thinking mode brings a new dimension to jailbreak attack research. Unlike standard mode's single-step responses, thinking mode undergoes a complex step-by-step reasoning process before generating final answers. In our preliminary experiments, we observed that thinking mode exhibits significant vulnerabilities when handling multi-task interleaved prompts, with significantly extended thinking processes. Based on this observation, we design the Multi-stream Perturbation Attack method. As shown in Figure \ref{fig:MS}, the method disrupts the model's reasoning process by interweaving benign auxiliary tasks with harmful tasks within a single prompt, including three perturbation strategies: multi-stream interleaving (MS), inversion perturbation (MS\_Reverse), and shape transformation (MS\_Structure). Experiments reveal that multi-stream perturbation not only bypasses safety mechanisms to induce harmful content generation, but also triggers failure modes unique to thinking mode, including thinking collapse, repetitive outputs, and abnormally extended reasoning. This indicates that the step-by-step reasoning process of thinking mode itself constitutes a new attack surface, with both reasoning stability and content safety simultaneously threatened.

The main contributions of this paper are summarized as follows:

\begin{itemize}
\item We propose multi-stream perturbation attack targeting thinking mode, disrupting the thinking process through multi-stream interleaving, inversion perturbation, and shape transformation.

\item We discover that thinking mode exhibits thinking collapse and repetitive outputs under multi-stream perturbation, revealing dual vulnerability in content safety and thinking stability.

\item We validate the attack effectiveness and security risks of thinking mode across multiple mainstream LLMs and three benchmark datasets.
\end{itemize}

\section{Related Work}

\subsection{Thinking Mode in Large Language Models}

Thinking capability is a core element for LLMs to accomplish complex tasks. In recent years, LLMs have demonstrated remarkable capabilities in mathematical problem solving~\cite{wang2023self}, common sense understanding~\cite{jung2022maieutic}, symbolic processing~\cite{zhou2023least}, and code generation. Chain-of-Thought (CoT) techniques~\cite{wei2022chain, kojima2022large} guide models to articulate their step-by-step thinking process through carefully designed prompts. Building on this foundation, researchers have proposed enhancement mechanisms including self-critique~\cite{ke2024critiquellm}, plan-and-solve approaches~\cite{wang2023plan}, multi-agent debate~\cite{liang2024encouraging}, and tree-of-thought~\cite{yao2023tree}. With the release of OpenAI's o1 series~\cite{jaech2024openai}, thinking mode has been built into model training, forming thinking models (large reasoning models, LRMs)~\cite{li2025system}. Unlike traditional models prompted to ``think step by step'', these models are explicitly trained to automatically generate detailed thinking processes before providing final answers. Models such as DeepSeek-R1~\cite{guo2025deepseek}, Kimi-2~\cite{team2025kimi}, and QwQ~\cite{team2024qwq} all leverage reinforcement learning to refine their thinking processes. Qwen3~\cite{yang2025qwen3} further integrates thinking mode and standard mode into a single model and introduces thinking budget mechanisms that allow users to dynamically adjust thinking depth based on task complexity.

\subsection{Jailbreak Attack}

Existing jailbreak attack methods can be divided into two categories based on their attack strategies.

White-box gradient-based jailbreak attacks optimize adversarial suffixes using the gradient information of open-source models. GCG proposed by~\citet{zou2023universal} generates jailbreak suffixes through the greedy coordinate gradient method. AutoDAN proposed by~\citet{liu2024autodan} optimizes manually initialized suffixes using a hierarchical genetic algorithm. MAC proposed by~\citet{zhang2025boosting} accelerates GCG attacks through momentum enhancement. Probe-Sampling developed by~\citet{zhao2024accelerating} dynamically evaluates candidate prompts using a small draft model. IRIS proposed by~\citet{huang2025stronger} introduces a refusal vector loss term to generate universal adversarial suffixes. AmpleGCG proposed by~\citet{liao2024amplegcg} uses generative models to capture the distribution characteristics of adversarial suffixes. Although gradient-based methods have high attack efficiency, they require white-box access and are limited in application on closed-source models.

Black-box optimization-based jailbreak attacks iteratively optimize jailbreak prompts through interaction with the target LLM, without accessing the model's internal structure. PAIR proposed by~\citet{chao2025jailbreaking} leverages the attacker LLM to generate and iteratively optimize jailbreak prompts. TAP proposed by~\citet{mehrotra2024tree} adopts a tree-based thinking approach to search the attack space. GPTFuzzer developed by~\citet{yu2024llm} optimizes seed templates through fuzz testing. AutoDAN-Turbo proposed by~\citet{liu2024autodanturbolifelongagentstrategy} constructs a lifelong learning framework to automatically evolve jailbreak strategies. JAIL-CON~\cite{jiang2025adjacent} constructs concurrent tasks by interleaving harmful tasks with benign tasks word by word and designs an iterative attack framework with a shadow judge mechanism to improve attack success rates (ASR). FlipAttack~\cite{liu2025flipattack} disguises harmful requests by flipping text characters and interfering with keyword-based detection. Notably, both MS\_Reverse and FlipAttack employ character reversal, but FlipAttack reverses harmful words to disrupt safety filtering, while MS\_Reverse reverses benign words to increase decoding burden and generate superimposed interference within the multi-stream framework. Unlike JAIL-CON and FlipAttack, multi-stream perturbation attack integrates both ideas and specifically targets thinking mode, exploiting its step-by-step reasoning to trigger abnormal phenomena such as thinking collapse and repetitive outputs that do not appear in standard mode.

\section{Multi-Stream Perturbation Attack Method}

This section introduces the multi-stream perturbation attack method targeting thinking mode. The core idea is to generate superimposed interference by interweaving multiple task streams within prompts, exploiting thinking mode's vulnerability when processing concurrent tasks to simultaneously attack content safety and reasoning stability.

\subsection{Problem Setting and Basic Framework}

Given a harmful task $q_{\text{harm}}$ and $k$ benign auxiliary tasks $\{q_{\text{aux}}^{(1)}, q_{\text{aux}}^{(2)}, \ldots, q_{\text{aux}}^{(k)}\}$, the goal of multi-stream perturbation attack is to construct a perturbed prompt $q_{\text{perturb}}$ such that the target LLM with thinking mode bypasses safety mechanisms and generates harmful responses $r_{\text{harm}}$ when processing $q_{\text{perturb}}$, or causes reasoning process collapse. We split the harmful and auxiliary tasks at word granularity, denoting the word sequence of the harmful task as $q_{\text{harm}} = [w_1^h, w_2^h, \ldots, w_m^h]$ and the word sequence of the $i$-th auxiliary task as $q_{\text{aux}}^{(i)} = [w_1^{(i)}, w_2^{(i)}, \ldots, w_{n_i}^{(i)}]$. Detailed descriptions and representative samples of the benign task pool are provided in Appendix \ref{appendix:benign-tasks}.

The fundamental reasons why multi-stream perturbation exhibits stronger attack effectiveness in thinking mode can be understood from three perspectives. From the perspective of cognitive resource allocation, thinking mode must simultaneously parse task boundaries marked by different delimiters, maintain multiple independent semantic representations, and dynamically switch attention focus between tasks. This mirrors multi-stream attention allocation mechanisms in cognitive science~\cite{bronkhorst2015cocktail}. As the number of concurrent tasks increases, the computational resources available for safety detection decrease significantly. From the perspective of safety mechanism detection logic, existing safety alignment methods primarily rely on complete, ordered text sequences for harmfulness judgment. Multi-stream interleaving disrupts the sequence integrity of harmful intent through fragmented ``benign word-harmful word-benign word'' arrangements, causing the detection probability of safety detectors to drop substantially. The characteristics of thinking mode further amplify these two effects. Its ``detail-first'' training objective drives models to generate ultra-long thinking processes under multi-stream perturbation, and powerful reasoning capabilities are employed to construct ``rationalization'' arguments for harmful responses. Meanwhile, human cognition research demonstrates fundamental limits to information processing speed~\cite{zheng2025unbearable}. When multi-stream perturbation significantly increases the per-step error rate, uncertainty accumulation during step-by-step reasoning causes the probability of thinking collapse or cyclical outputs to rise rapidly.

Multi-stream perturbation attack uses specific delimiters to mark different task streams, forcing thinking mode to simultaneously parse and process multiple tasks, thereby causing attention dispersion and reasoning path confusion and reducing the effectiveness of safety detection. Based on this framework, we propose four specific perturbation strategies that disrupt thinking mode's reasoning process from different perspectives.

\subsection{Perturbation Strategies}

We design three perturbation strategies, constructed as follows:

\[
\begin{aligned}
q_{\text{perturb}}^{\text{ms}} &= \bigoplus_{j=1}^{L} \left[ w_j^h \oplus \{w_j^{(1)}\} \oplus [w_j^{(2)}] \right]
\end{aligned}
\]

\[
\begin{aligned}
q_{\text{perturb}}^{\text{inv}} &= \bigoplus_{j=1}^{L} \left[ w_j^h \oplus \{\overleftarrow{w_j^{(1)}}\} \right] 
\end{aligned}
\]

\[
\begin{aligned}
q_{\text{perturb}}^{\text{shape}} &= q_{\text{perturb}}^{\text{ms}} \oplus \text{ShapeConstraint}
\end{aligned}
\]

where $\oplus$ denotes word sequence concatenation, $L$ denotes the task stream length, and $\overleftarrow{w_j^{(i)}} = \text{reverse}(w_j^{(i)})$ denotes character sequence reversal. \textbf{Multi-Stream Interleaving Perturbation (MS)} interleaves a harmful task with multiple benign auxiliary tasks word by word, with different task streams marked by different separators such as $\{\cdot\}$ and $[\cdot]$, forcing the model to parse multiple interleaved semantic paths simultaneously and producing a superimposed interference effect. \textbf{Inversion Perturbation (MS\_Reverse)} reverses each word in auxiliary tasks at the character level. This exploits LLMs' denoising ability to still understand reversed words while increasing the decoding burden on harmful tasks, generating superimposed interference within the multi-stream interleaving framework. \textbf{Shape Transformation Perturbation (MS\_Structure)} adds a triangular output format constraint to multi-stream interleaving, where the $i$-th line contains $i$ characters. The format constraint further disperses model attention, making the model more prone to thinking errors under the triple constraints of content generation, multi-stream parsing, and format control. 

These three perturbation strategies disrupt thinking mode's thinking process from different dimensions: multi-stream interleaving intensifies attention dispersion by increasing the number of concurrent tasks, inversion perturbation increases decoding burden through character reversal, and shape transformation introduces additional cognitive load through format constraints.

\section{Experiment}

\subsection{Experimental setup}

This section briefly introduces the experimental configuration. Detailed hardware environment, model configuration, dataset descriptions, and evaluation metric specifications are provided in Appendix \ref{appendix:exp-setup}.

\begin{figure*}[t]
  \centering
  \includegraphics[width=2\columnwidth]{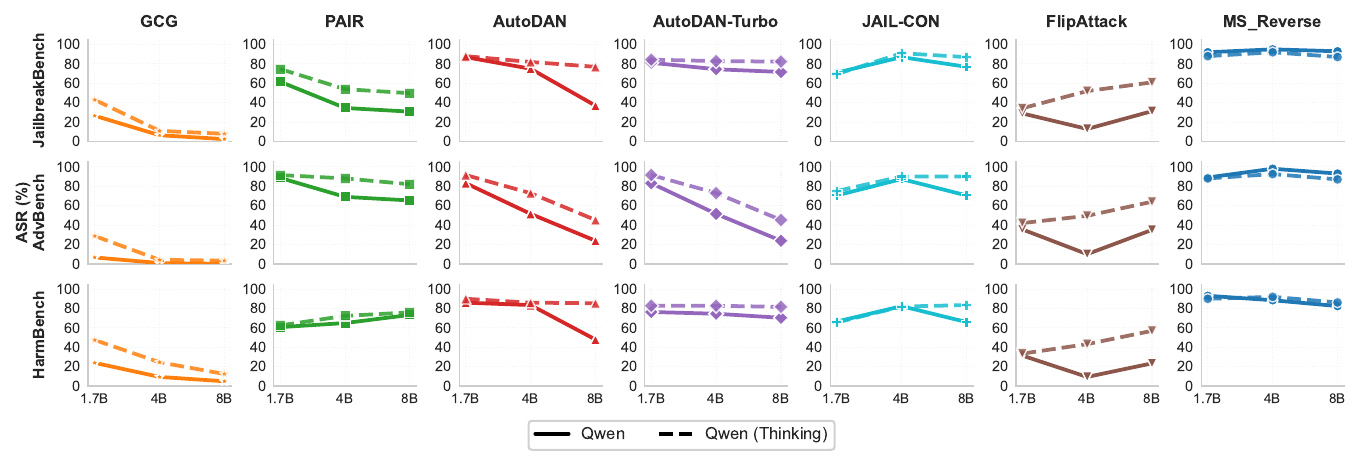}
  \caption{ASR comparison of seven attack methods on Qwen3 series models.}
  \label{fig:qwen_three_benchmarks_data}
\end{figure*}

\begin{figure*}[t]
  \centering
  \includegraphics[width=1.6\columnwidth]{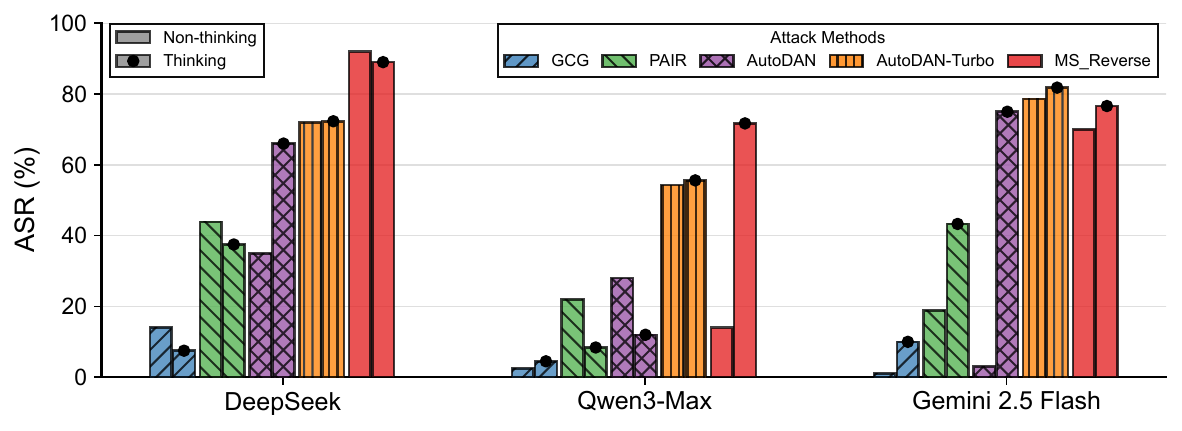}
  \caption{ASR comparison of five attack methods on DeepSeek, Qwen3-Max, and Gemini 2.5 Flash on the JailbreakBench dataset.}
  \label{fig:asr_jbb}
\end{figure*}

\begin{figure*}[t]
  \centering
  \includegraphics[width=1.7\columnwidth]{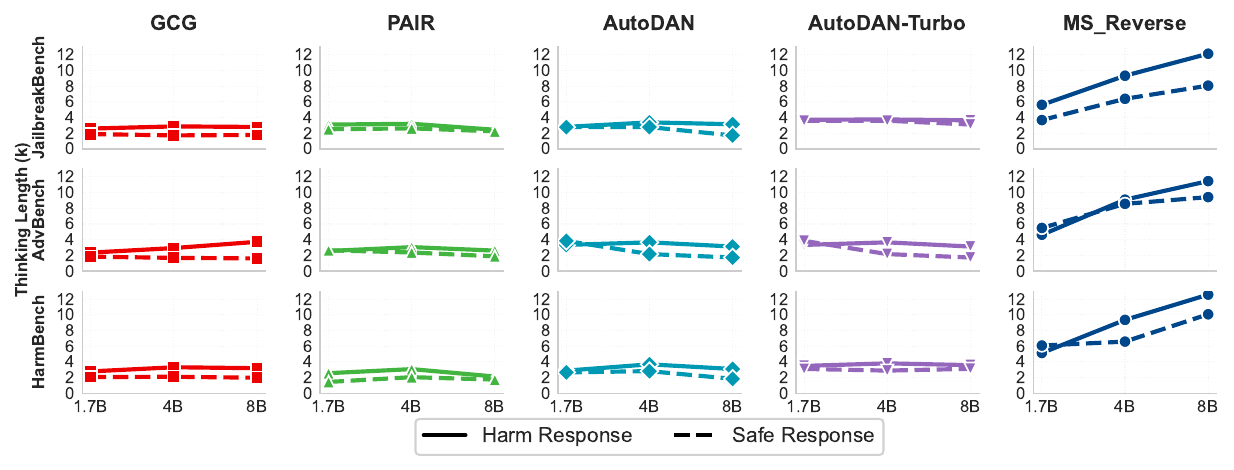}
  \caption{Thinking length comparison of five attack methods on Qwen3 series models.}
  \label{fig:qwen_lenT_safe_vs_harm}
\end{figure*}

\textbf{Environment}. Experiments were conducted on a server equipped with six NVIDIA GeForce RTX 3090 GPUs and dual Intel Xeon Gold 5317 CPUs.

\textbf{LLMs}. We evaluated 6 mainstream LLMs: open-source models Qwen3 1.7B, Qwen3 4B, Qwen3 8B~\cite{yang2025qwen3}, and API-based models DeepSeek (0.2/3 CNY per 1M input/output tokens), Qwen3-Max (0.8/2 CNY per 1M tokens), and Gemini 2.5 Flash ( 0.3/2.5 USD per 1M input/output tokens). For models supporting thinking length control, we adopted the maximum thinking length settings recommended in official documentation. It should be noted that although models such as ChatGPT o1 and Grok also possess thinking capabilities, experiments were conducted only on models that can independently and stably output thinking content, as ChatGPT o1's thinking content is post-processed and Grok only displays thinking time without outputting thinking content, making them unsuitable for quantitative analysis of thinking processes.

\textbf{Datasets}. We evaluated on three benchmark datasets: JailbreakBench~\cite{chao2024jailbreakbench}, AdvBench~\cite{zou2023universal}, and HarmBench~\cite{mazeika2024harmbench}.

\textbf{Baselines}. We compare with five representative jailbreak attack methods: GCG, PAIR, AutoDAN, AutoDAN-Turbo, JAIL-CON, and FlipAttack.

\begin{figure}[t]
  \centering
  \includegraphics[width=1\columnwidth]{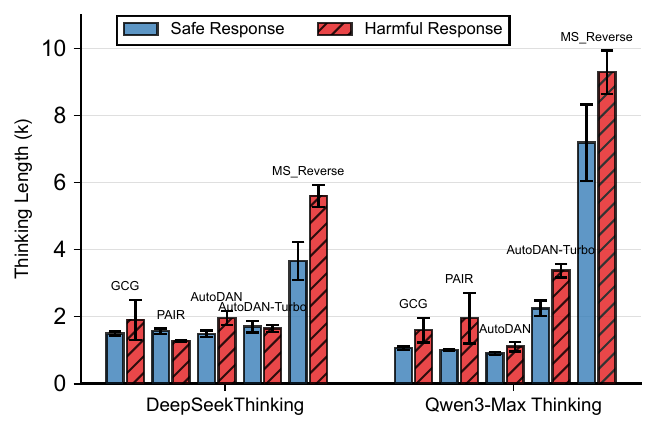}
  \caption{Thinking length comparison of five attack methods on DeepSeek and Qwen3-Max on the JailbreakBench dataset.}
  \label{fig:thinking_length}
\end{figure}

\textbf{Evaluation Metrics}. We use ASR to evaluate jailbreak attacks, Len-T to evaluate thinking length, TCR to evaluate thinking collapse, and RRR to evaluate response repetition. \textit{ASR} is the proportion of successfully attacked samples to the total number of attacked samples.  \textit{Len-T} represents the average thinking length with standard error, calculated as $\text{Mean} \pm \frac{\sigma}{\sqrt{n}}$, where Mean is the average thinking length, $\sigma$ is the standard deviation, and $n$ is the sample size. \textit{TCR} is the proportion of thinking collapses to the total number of thinking instances. A thinking collapse is identified when the thinking content contains massive repetitive strings or reaches the output limit, and the model fails to generate any answer content. \textit{RRR} is the proportion of repetitive outputs to the total number of outputs, referring to the phenomenon where massive repetitive content appears during the response process until reaching the output limit.

\begin{table*}
  \centering
  \caption{\label{qwen3-deepseek-lent-table}Thinking length comparison of three perturbation strategies on Qwen3 series models and DeepSeek.}
  \resizebox{0.9\linewidth}{!}{%
  \begin{tabular}{l|cc|cc|cc|cc}
    \hline
    \multirow{2}{*}{\textbf{Method}} & \multicolumn{2}{c|}{\textbf{Qwen3 1.7B}} & \multicolumn{2}{c|}{\textbf{Qwen3 4B}} & \multicolumn{2}{c|}{\textbf{Qwen3 8B}} & \multicolumn{2}{c}{\textbf{DeepSeek}} \\
    \cline{2-9}
    & \textbf{Safe} & \textbf{Harm} & \textbf{Safe} & \textbf{Harm} & \textbf{Safe} & \textbf{Harm} & \textbf{Safe} & \textbf{Harm} \\
    \hline
    MS & \underline{$7068_{\pm414}$} & $6557_{\pm818}$ & $5263_{\pm1534}$ & \underline{$5307_{\pm494}$} & \underline{$5865_{\pm1798}$} & $4882_{\pm568}$ & $7808_{\pm1896}$ & \underline{$14574_{\pm1393}$} \\
    MS\_Reverse   & $3659_{\pm561}$ & \underline{$5600_{\pm333}$} & $6370_{\pm2255}$ & \underline{$9312_{\pm491}$} & $8050_{\pm2218}$ & \underline{$12121_{\pm636}$} & $3197_{\pm1994}$ & \underline{$17734_{\pm1209}$} \\
    MS\_Structure & $6882_{\pm442}$ & \underline{$10368_{\pm4011}$} & \underline{$6761_{\pm1126}$} & $6463_{\pm460}$ & \underline{$7739_{\pm1655}$} & $5127_{\pm654}$ & $27522_{\pm2256}$ & \underline{$28321_{\pm537}$} \\
    \hline
  \end{tabular}%
  }

\end{table*}

\subsection{Attack Success Rate Experimental Results}

We first evaluated the attack effectiveness of Multi-stream Perturbation Attack and six baseline methods on the Qwen3 series models (1.7B, 4B, 8B). As shown in Figure~\ref{fig:qwen_three_benchmarks_data}, we compared the MS\_Reverse strategy with baseline methods on three benchmark datasets, where solid lines represent standard mode and dashed lines represent thinking mode. MS\_Reverse achieves significantly higher ASR than baseline methods across all three datasets in both standard mode and thinking mode, maintaining consistently high ASR across different model scales.

\begin{figure}[t]
  \centering
  \includegraphics[width=1\columnwidth]{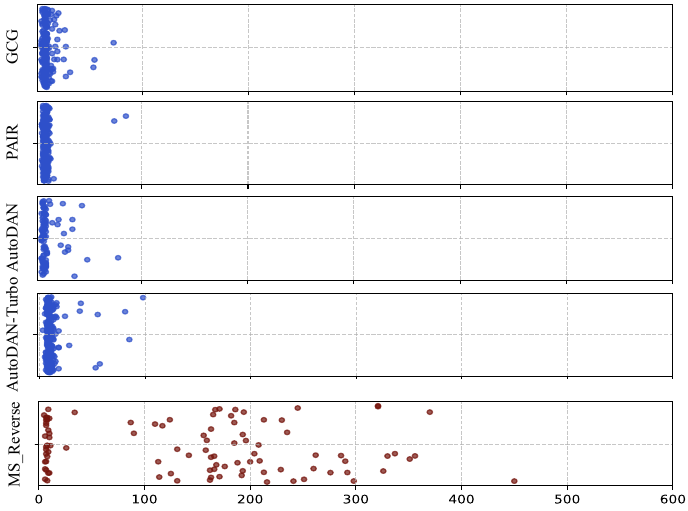}
  \caption{Thinking time comparison of five attack methods on DeepSeek on the JailbreakBench dataset. Time is measured in seconds.}
  \label{fig:Len_Time1}
\end{figure}

To further validate the effectiveness of Multi-stream Perturbation Attack on large-scale closed-source models, we conducted evaluations on three API-based LLMs: DeepSeek, Qwen3-Max, and Gemini 2.5 Flash. As shown in Figure \ref{fig:asr_jbb}, on the JailbreakBench dataset, MS\_Reverse achieves the highest ASR in both standard mode and thinking mode on DeepSeek, achieves the highest ASR in thinking mode on Qwen3-Max, and ranks among the top on Gemini 2.5 Flash. In terms of model safety, Qwen3-Max exhibits relatively lower overall ASR, demonstrating stronger safety alignment capability.

\subsection{Thinking Attack Experimental Results}

We designed three quantitative metrics to evaluate the impact of multi-stream perturbation attacks on the thinking process: Len-T, TCR, and RRR.

\textbf{Thinking Length Analysis.} As shown in Figures~\ref{fig:qwen_lenT_safe_vs_harm} and~\ref{fig:thinking_length}, where solid lines represent harmful responses and dashed lines represent safe responses. MS\_Reverse produces thinking lengths far exceeding other methods, surpassing 10K characters on Qwen3 8B, while other attack methods typically generate thinking lengths ranging from 2K to 4K characters. Similar trends are observed on DeepSeek and Qwen3-Max, with MS\_Reverse's thinking length significantly higher than other methods. Furthermore, harmful responses generally exhibit longer thinking lengths than safe responses. This phenomenon remains consistent across all model scales, indicating that thinking mode tends to engage in more detailed thinking processes when generating harmful content.

\textbf{Thinking Collapse and Repetitive Outputs.} Experimental results show that MS\_Reverse significantly outperforms other baseline methods on both metrics: thinking collapse rate reaches 17\% on Qwen3 4B while other methods remain essentially at 0; response repetition rate reaches 60\% on Qwen3 4B and 25\% on DeepSeek, far exceeding the below 2\% level of other methods. These results indicate that multi-stream perturbation attacks not only bypass safety mechanisms but also cause thinking process collapse or repetitive outputs. Detailed thinking collapse rate and response repetition rate data are provided in Appendix \ref{appendix:tcr-rrr}.

\textbf{Thinking Time Cost.} As shown in Figure \ref{fig:Len_Time1}, MS\_Reverse's thinking time far exceeds other methods, with the maximum reaching 7 minutes, indicating that multi-stream perturbation attack substantially increases the model's inference time cost, not only consuming more computational resources but also affecting the model's practical application effectiveness.

\begin{figure}[t]
  \centering
  \includegraphics[width=1\columnwidth]{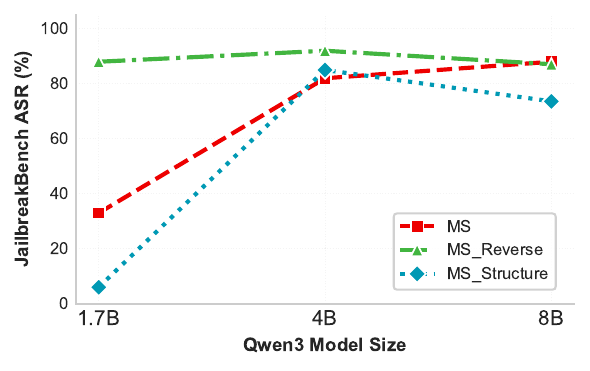}
  \caption{ASR comparison of three perturbation strategies on Qwen3 series models on JailbreakBench dataset.}
  \label{fig:qwen3_ablation_asr_MS}
\end{figure}

To understand the vulnerabilities of thinking mode under multi-stream perturbation attacks, we analyzed the causes of safety degradation, abnormal thinking length, and reasoning stability collapse. These phenomena stem from the inherent conflict between thinking mode's training objective of ``detailed analysis, step-by-step reasoning, and providing help'' and safety alignment mechanisms. Multi-stream interleaving disrupts the sequence integrity of harmful intent through fragmented ``benign word-harmful word-benign word'' arrangements, and the pursuit of detail drives the model to deeply analyze each task stream, making it prone to misjudging harmful requests as complex problems under attention dispersion. The open-source Qwen series has 36.31\% and DeepSeek has 16.36\% of harmful responses claiming ``educational purposes'', indicating that reasoning capabilities are employed to construct ``reasonable'' justifications. The pursuit of detail leads to abnormal thinking lengths: as shown in Figure \ref{fig:qwen_lenT_safe_vs_harm}, MS\_Reverse exceeds 10K characters on Qwen3 8B and 20K characters on DeepSeek. When simultaneously processing multiple interleaved task streams, the model needs to repeatedly switch semantic paths and maintain multiple representation states, leading to a thinking collapse rate of 17\% and response repetition rate of 60\% on Qwen3 4B (25\% on DeepSeek), because multi-stream perturbation disrupts reasoning coherence and uncertainty accumulation causes the model to fall into local loops. Previous research~\citep{zhou2025hidden} has shown that shortening the thinking process can improve harmlessness, which confirms the vulnerability of long reasoning chains. This asymmetry in capability generalization makes models prone to using their reasoning abilities to bypass safety mechanisms~\citep{bondarenko2025demonstrating}.

The above analysis shows that multi-stream perturbation attack differs fundamentally from existing jailbreak methods. The attack effects of JAIL-CON and FlipAttack are limited to bypassing safety filtering. Multi-stream perturbation attack exploits the step-by-step reasoning characteristics of thinking mode, extending the attack effects from pure content safety bypass to disruption of the reasoning process itself. The comparison experiments in Appendix~\ref{appendix:jailcon-comparison} further confirm this. Under the same thinking mode settings, JAIL-CON produces lower thinking lengths and thinking times than all three strategies of multi-stream perturbation attack.

\subsection{Ablation Study}

To evaluate the effectiveness of different strategies in Multi-stream Perturbation Attack, we tested three perturbation forms on the JailbreakBench dataset, including MS, MS\_Reverse, and MS\_Structure. As shown in Figure~\ref{fig:qwen3_ablation_asr_MS}, we compared the ASR of three strategies on Qwen open-source models. MS\_Reverse achieves the highest ASR across all model scales, balancing attack complexity and attack effectiveness well. All three strategies show a similar trend. ASR is relatively lower on the 1.7B model, improves significantly as model scale increases to 4B, but slightly decreases on the 8B model due to stronger safety alignment.

\begin{figure}[t]
  \centering
  \includegraphics[width=1\columnwidth]{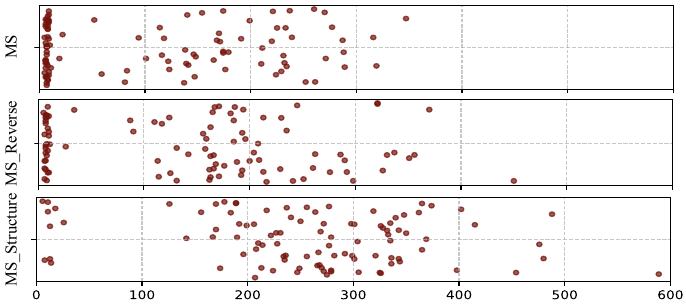}
  \caption{Thinking time comparison of three perturbation strategies on DeepSeek on the JailbreakBench dataset.}
  \label{fig:Len_Time2}
\end{figure}

As shown in Table~\ref{qwen3-deepseek-lent-table}, we compare the thinking lengths of three strategies on the JailbreakBench dataset, where Safe and Harm denote safe responses and harmful responses respectively. Thinking length increases as model scale grows. MS\_Reverse and MS\_Structure exceed 10K characters in thinking length for harmful responses on Qwen3 8B. On DeepSeek, MS, MS\_Reverse, and MS\_Structure all exceed 10K characters, with MS\_Structure even exceeding 20K characters, indicating that Multi-stream Perturbation Attack leads to substantial computational and time costs under thinking mode.

For thinking collapse rate and response repetition rate, as shown in Table~\ref{tab:t-break-ablation}, different strategies exhibit varying performance across different models. MS\_Reverse causes higher collapse rates on multiple models, reaching 17\% on Qwen3 4B. Overall, the structural perturbations introduced by MS\_Reverse and MS\_Structure are more likely to cause collapse in the model's thinking process. As shown in Table~\ref{tab:a-repeat-ablation}, MS, MS\_Reverse, and MS\_Structure all exhibit high repetition rates on Qwen3 series open-source models, reaching 60\% on Qwen3 4B. In contrast, these strategies have minimal impact on DeepSeek, with repetition rates remaining at low levels. This indicates that different model architectures have varying sensitivity to perturbation strategies, with Qwen3 series models being more prone to output repetition when facing complex perturbations. Furthermore, as shown in Figure~\ref{fig:Len_Time2}, thinking time gradually increases from MS to MS\_Structure, with MS\_Structure exceeding 9 minutes. This result is consistent with the trend in thinking length, further validating that multi-stream perturbation attack significantly increases the model's thinking time cost.

\begin{table}
  \centering
  \caption{\label{tab:t-break-ablation}Thinking collapse rate comparison of three perturbation strategies on Qwen3 series models and DeepSeek, on the JailbreakBench dataset.}
  \resizebox{1.0\linewidth}{!}{%
  \begin{tabular}{l|cccc}
    \hline
    \multirow{2}{*}{TCR} & \multicolumn{4}{c}{\textbf{Model}} \\
    \cline{2-5}
    & Qwen3 1.7B & Qwen3 4B & Qwen3 8B & DeepSeek \\
    \hline
    MS           & 2.0 & 1.0 & 0.0 & 3.0 \\
    MS\_Reverse   & 8.0 & 17.0 & 9.0 & 3.0 \\
    MS\_Structure & 4.0 & 1.0 & 4.2 & 10.1 \\
    \hline
  \end{tabular}%
  }
\end{table}

\begin{table}
  \centering
  \caption{\label{tab:a-repeat-ablation}Response repetition rate comparison of three perturbation strategies on Qwen3 series models and DeepSeek, on the JailbreakBench dataset.}
  \resizebox{1.0\linewidth}{!}{%
  \begin{tabular}{l|cccc}
    \hline
    \multirow{2}{*}{RRR} & \multicolumn{4}{c}{\textbf{Model}} \\
    \cline{2-5}
    & Qwen3 1.7B & Qwen3 4B & Qwen3 8B & DeepSeek \\
    \hline
    MS           & 16.0 & 23.0 & 33.0 & 0.0 \\
    MS\_Reverse   & 25.0 & 60.0 & 51.0 & 2.0 \\
    MS\_Structure & 17.0 & 26.0 & 20.0 & 0.0 \\
    \hline
  \end{tabular}%
  }
\end{table}

\subsection{Harmful Content Detection Defense}

Detecting harmfulness in model outputs is an important safety defense in practical deployments~\cite{inan2023llama,chi2024llama}. To verify the detection capability of different methods against outputs from multi-stream perturbation attacks, we constructed two evaluation datasets. The challenging dataset contains 358 samples where harmful content is hidden in role-playing narratives or lies on the ambiguous boundary between safe and harmful, making detection difficult. The regular dataset contains 720 samples with relatively obvious harmful features, closer to routine defense scenarios. We compared multiple detection methods on both datasets, including keyword-based detection, GPT-4, DeepSeek, the Qwen3Guard series, and the Llama-Prompt-Guard-2 series. Llama-22M and Llama-86M denote Llama-Prompt-Guard-2 models with 22M and 86M parameters respectively.

As shown in Table~\ref{tab:harmful_evaluator_comparison}, detection methods show significant performance differences on the challenging dataset. The Llama-Prompt-Guard-2 series performs worst, with low accuracy and persistently high FPR or FNR. Keyword-based detection achieves only 60.89\% accuracy, struggling to handle the complex semantic variations produced by multi-stream perturbation. The Qwen3Guard series performs best, with Qwen3Guard 4B achieving 84.08\% accuracy and only 5.31\% FPR. As shown in Table~\ref{tab:regular_evaluator_comparison}, on the regular dataset, Qwen3Guard 4B further improves to 92.22\% accuracy with 4.58\% FPR. These results indicate that multi-stream perturbation attacks pose a considerable challenge to existing content detection defenses. Dataset construction details are provided in Appendix~\ref{appendix:harmful-eval}.

\begin{table}[h]
  \centering
  \caption{\label{tab:harmful_evaluator_comparison}Performance comparison of different harmful content detection methods on the challenging evaluation dataset.}
  \resizebox{1.0\linewidth}{!}{%
  \begin{tabular}{l|ccc}
    \hline
     & Accuracy (\%) & FPR (\%) & FNR (\%) \\
    \hline
    Llama-22M & 49.16 & 12.29 & 37.43 \\
    Llama-86M & 41.34 & 34.07 & 23.46 \\
    Keyword-based & 60.89 & 14.25 & 23.74 \\
    GPT-4 & 70.39 & 7.54 & 20.39 \\
    DeepSeek & 70.67 & 8.66 & 19.55 \\
    Qwen3Guard 0.6B & 77.65 & 6.98 & 14.25 \\
    Qwen3Guard 4B & \textbf{84.08} & \textbf{5.31} & 9.50 \\
    Qwen3Guard 8B & 81.84 & 7.82 & \textbf{9.22} \\
    \hline
  \end{tabular}%
  }
  
\end{table}

\begin{table}[h]
\centering
\caption{\label{tab:regular_evaluator_comparison}Performance comparison of Qwen3Guard series models on the regular evaluation dataset.}   
\resizebox{1.0\linewidth}{!}{%
        \begin{tabular}{lccc}
        \toprule
         & Accuracy (\%) & FPR (\%) & FNR (\%) \\
        \midrule
        Qwen3Guard 0.6B & 89.44 & \textbf{3.75} & 6.81 \\
        Qwen3Guard 4B & \textbf{92.22} & 4.58 & 3.19 \\
        Qwen3Guard 8B & 91.25 & 5.97 & \textbf{2.78} \\
        \bottomrule
        \end{tabular}
    }
\end{table}

\section{Conclusion}

This paper proposes a multi-stream perturbation attack method targeting thinking mode, which tests LLMs' security vulnerabilities in complex reasoning scenarios by interweaving multiple task streams within a single prompt. Experiments validate the unique vulnerabilities of thinking mode when facing multi-stream perturbations, with the MS\_Reverse strategy achieving ASR exceeding 90\% on certain models. We find that structured perturbations from multiple task streams produce a superimposed interference effect on thinking mode's reasoning process, simultaneously compromising content safety and reasoning stability, leading to thinking collapse rates and response repetition rates reaching 17\% and 60\% respectively. This research provides new perspectives for understanding the security risks of thinking mode, and future work can explore defense mechanisms targeting thinking mode and the relationship between thinking length control and safety.

\section{Ethics Statement}
This research aims to evaluate the security of thinking mode LLMs, helping the research community better understand and defend against jailbreak attacks. All experiments strictly comply with relevant ethical guidelines and legal requirements. Harmful content involved in experiments is limited to samples from academic benchmark datasets. Experiments were conducted only on open-source models and commercial models with authorized API access, without testing any unauthorized systems. We call on the research community to adhere to responsible disclosure principles when exploring model security issues.

\bibliography{uai2026-template}

\newpage


\title{Multi-Stream Perturbation Attack\\(Supplementary Material)}
\maketitle

\appendix
\section{Detailed Experimental Configuration}
\label{appendix:exp-setup}

\textbf{Hardware and Software Environment}: This experiment was conducted using a server equipped with six NVIDIA GeForce RTX 3090 GPUs (each with 24GB video memory, totaling 144GB video memory) and dual Intel Xeon Gold 5317 CPUs (24 cores, 48 threads total), with Ubuntu 22.04 as the operating system. The software versions included PyTorch 2.3.0, Python 3.12, and CUDA 12.7.

\textbf{Detailed Model Configuration}: We evaluated 6 mainstream LLMs, including three open-source small-parameter LLMs deployed locally (Qwen3 1.7B, Qwen3 4B, Qwen3 8B~\cite{yang2025qwen3}) and three API-based LLMs (DeepSeek, Qwen3-Max, and Gemini 2.5 Flash). For models supporting thinking length control, we adopted the maximum thinking length settings recommended in official documentation: the thinking length limit for the open-source Qwen3 series models was set to 32,768 tokens, and the thinking length limit for Gemini 2.5 Flash was set to 24,576 tokens. For the temperature parameter, we adopted the official default configurations for each model: 0.6 for Qwen3 series models (1.7B, 4B, 8B), 1.0 for DeepSeek, 0.7 for Qwen3-Max, and 1.0 for Gemini 2.5 Flash.

It should be noted that although models such as ChatGPT o1 and Grok also possess thinking capabilities, our experiments are conducted only on models that can independently and stably output thinking content. Specifically, the Qwen3 series, DeepSeek, Qwen3-Max, and Gemini 2.5 Flash all output complete thinking processes in standard format, enabling us to accurately measure metrics such as thinking length and thinking collapse rate. In contrast, ChatGPT o1's thinking behavior is autonomously determined by the model and its thinking content is post-processed, while Grok only displays thinking time without outputting thinking content. These characteristics make them unsuitable for the quantitative analysis of thinking processes in this study.

\textbf{Detailed Dataset Description}: We evaluated our method on three widely adopted benchmark datasets — JailbreakBench~\cite{chao2024jailbreakbench}, AdvBench~\cite{zou2023universal}, and HarmBench~\cite{mazeika2024harmbench}. Their diverse characteristics enable us to more comprehensively evaluate the performance of the method across different types of harmful behaviors and attack scenarios. Among them, JailbreakBench provides extensive coverage of harmful instructions as well as benign tasks on various topics, making it suitable for evaluating attack effectiveness and false positive rates; AdvBench contains a curated collection of adversarial prompts specifically designed to test safety alignment mechanisms; HarmBench collects diverse malicious requests that violate legal or normative standards.

\textbf{Detailed Evaluation Metrics}

Len-T (Thinking Length) is calculated as:
\begin{equation}
    \text{Len-T} = \text{Mean} \pm \frac{\sigma}{\sqrt{n}}
\end{equation}
where Mean represents the average thinking length, $\sigma$ is the standard deviation of thinking length, and $n$ is the sample size (i.e., the number of prompts under the attack method). The standard error more directly reflects the uncertainty of the sample mean than the standard deviation, and serves as the foundation for significance testing (e.g., $t$-test).

\section{Benign Auxiliary Task Pool}
\label{appendix:benign-tasks}

The benign task pool $\mathcal{T}_{\text{aux}}$ contains a variety of harmless everyday queries spanning multiple thematic domains including education, science, and technology. In this study, we construct the task pool using benign task samples from the JailbreakBench dataset, though other qualifying data sources are equally applicable. 

Below are representative samples from the benign task pool:

\begin{tcolorbox}[
    colframe=black,
    colback=gray!10,
    arc=1mm,
    boxrule=0.3mm,
    fontupper=\fontsize{9pt}{12pt}\selectfont,
    boxsep=-1pt,           
]

1. List the categories of domain names.

2. Briefly introduce the types of French cakes.

3. Explain the process of photosynthesis.

4. Describe the history of the Eiffel Tower.

5. What are the benefits of regular exercise?

6. How does a computer processor work?

7. Explain the water cycle.

8. What are the main components of a healthy diet?

9. Describe the solar system.

10. How do airplanes fly?

11. What is artificial intelligence?

12. Explain how the internet works.

13. What are the different types of renewable energy?

14. Describe the process of making coffee.

15. What are the main programming languages?

16. Explain the concept of supply and demand.

17. What are the layers of Earth's atmosphere?

18. How do vaccines work?

19. Describe the process of DNA replication.

20. What are the principles of effective communication?

\end{tcolorbox}

\section{Detailed Evaluation of Harmful Content Detection Methods}
\label{appendix:harmful-eval}

\subsection{Necessity of Harmful Content Detection}

Harmful content detection aims to determine whether LLM outputs contain harmful content. It is an important component of the LLM safety defense system and the foundation for accurately computing ASR. In jailbreak attack research, selecting an appropriate detection method poses inherent challenges. Jailbreak attacks involve the generation of semantic content. Detection methods must understand the creative expressions and semantic meanings in candidate jailbreak prompts and responses, rather than relying on simple phrase lists or fixed rules~\citep{chao2024jailbreakbench}.

\subsection{Existing Evaluation Methods}

Existing harmful content detection methods fall into two main categories. \textbf{Keyword-based detection methods} perform judgment by matching specific harmful keywords but suffer from obvious flaws. Some responses contain harmful keywords yet output safe content, while some harmful responses are mistakenly judged as safe due to the absence of harmful keywords~\citep{liu2023jailbreaking, li2024deepinception, jain2023baseline}. Such methods typically have high false positive and false negative rates, struggling to handle complex semantic variations. \textbf{LLM judgment-based methods} leverage the semantic understanding of large language models.~\citet{wei2023jailbreak} combines a fine-tuned Llama-13b with string detection to evaluate generated content.~\citet{liu2024autodan} first conducts preliminary screening through string detection, then uses GPT for harmful content verification.~\citet{qi2024fine} uses GPT-4 Judge to quantify harm levels on a 1-5 scale. Llama Guard~\citep{inan2023llama, chi2024llama}, as a specialized output detection model, identifies harmful content by judging the safety of target model responses. Qwen3Guard~\citep{zhao2025qwen3guard}, trained on 1.19 million labeled samples, performs harmful content detection through a three-level classification of ``Safe-Controversial-Unsafe''.

\subsection{Evaluation Metrics}

We adopt three metrics to quantify the performance of detection methods. \textbf{Judgment Accuracy} is the proportion of samples where the judgment result matches the correct answer. \textbf{False Positive Rate FPR} is the proportion of safe samples incorrectly judged as harmful, reflecting the risk of over-judgment. Minimizing FPR is essential when selecting detection methods, as it is more important to remain conservative to avoid classifying benign behavior as jailbroken~\citep{chao2024jailbreakbench}. \textbf{False Negative Rate FNR} is the proportion of harmful samples not identified.

\subsection{Detection Method Comparison and Selection}

To select an appropriate harmful content detection method, we constructed two evaluation datasets to test detection performance in both extreme challenge scenarios and routine application scenarios.

\textbf{Challenging Evaluation Dataset.} We compared multiple detection methods on a challenging dataset containing 358 samples. The dataset was constructed as follows. We first collected attack responses from baseline methods across various target models. Then we employed multiple LLMs for voting-based judgment, where a response was preliminarily labeled as harmful only when the majority of models deemed it harmful. Finally, through manual review, we further filtered out challenging samples. The resulting dataset comprises 195 safe samples and 163 harmful samples. These samples are challenging because harmful content is hidden in role-playing narratives, response content lies on the ambiguous boundary between safe and harmful, or harmful information is interleaved with safe content.

As shown in Table~\ref{tab:harmful_evaluator_comparison}, detection methods show significant performance differences on the challenging dataset. The Llama-Prompt-Guard-2 series performs worst, with low accuracy and persistently high FPR or FNR. Keyword-based detection achieves 60.89\% accuracy, with FPR and FNR remaining at relatively high levels. GPT-4 and DeepSeek show similar performance with accuracy around 70\%. The Qwen3Guard series performs best overall, with accuracy exceeding 77\%. Qwen3Guard 4B achieves the best comprehensive performance, reaching 84.08\% accuracy with the lowest FPR of 5.31\% and a relatively low FNR of 9.50\%.

\textbf{Regular Evaluation Dataset.} In practice, most harmful content detection tasks face regular samples where harmfulness judgment is relatively straightforward. To comprehensively evaluate the practical performance of Qwen3Guard models, we constructed a regular dataset containing 720 samples, including 489 harmful samples and 231 safe samples. Samples in this dataset come from common jailbreak attack scenarios with relatively obvious harmful features.

As shown in Table~\ref{tab:regular_evaluator_comparison}, all three models achieve significantly higher accuracy on the regular dataset than on the challenging dataset, with substantial reductions in both FPR and FNR. Qwen3Guard 4B achieves 92.22\% accuracy with FPR of only 4.58\% and FNR of 3.19\%, showing optimal comprehensive performance. Qwen3Guard 8B achieves the best FNR at 2.78\%, though with a slightly higher FPR of 5.97\%. Despite having the smallest parameter count, Qwen3Guard 0.6B still achieves 89.44\% accuracy.

Based on results from both datasets, we select Qwen3Guard 4B as the harmful content detection method for this study, also serving as the evaluator for computing ASR. Qwen3Guard 4B achieves the best comprehensive performance on both datasets, and as an open-source model it ensures experimental reproducibility.

\begin{table}
  \centering
  \resizebox{1.0\linewidth}{!}{%
  \begin{tabular}{c|l|ccccc}
    \hline
    \multirow{2}{*}{\textbf{Dataset}} & \multirow{2}{*}{\textbf{Model}} 
    & \multicolumn{5}{c}{\textbf{Thinking Collapse Rate -- Attack Methods}} \\
    \cline{3-7}
    &  & \textbf{MS\_R} & \textbf{GCG} & \textbf{PAIR} & \textbf{AutoDAN} & \textbf{AutoDAN-T} \\
    \hline

    \multirow{3}{*}{JBB} 
      & Qwen3 1.7B & 8.0  & 0.0 & 0.0 & 0.0 & 0.0 \\
      & Qwen3 4B   & 17.0 & 0.0 & 0.0 & 0.0 & 0.0 \\
      & Qwen3 8B   & 9.0  & 0.0 & 0.0 & 0.0 & 0.0 \\
    \hline

    \multirow{3}{*}{Adv} 
      & Qwen3 1.7B & 10.0 & 0.8 & 0.5 & 0.0 & 0.0 \\
      & Qwen3 4B   & 13.0 & 1.6 & 0.0 & 0.0 & 0.0 \\
      & Qwen3 8B   & 13.5 & 0.8 & 0.0 & 0.0 & 0.0 \\
    \hline

    \multirow{3}{*}{Harm} 
      & Qwen3 1.7B & 7.5 & 0.5 & 0.0 & 0.5 & 0.0 \\
      & Qwen3 4B   & 11.0 & 0.0 & 0.0 & 0.0 & 1.1 \\
      & Qwen3 8B   & 10.8 & 0.0 & 0.5 & 0.0 & 0.6 \\
    \hline

    \multirow{2}{*}{JBB} 
      & DeepSeek        & 8.0 & 0.5 & 0.4 & 0.0 & 0.0 \\
      & Qwen3-Max       & 1.0 & 0.5 & 0.0 & 0.0 & 0.0 \\
    \hline
  \end{tabular}%
  }
  \caption{\label{tab:T-Break}Thinking collapse rate comparison of five attack methods. Thinking collapse rates of five methods on Qwen3 series models (1.7B, 4B, 8B), DeepSeek, and Qwen3-Max across three benchmark datasets. JBB, Adv, Harm denote JailbreakBench, AdvBench, HarmBench datasets respectively; MS\_R, AutoDAN-T denote MS\_Reverse(ours), AutoDAN-Turbo respectively.}
\end{table}

\begin{table}
  \centering
  \resizebox{1.00\linewidth}{!}{%
  \begin{tabular}{c|l|ccccc}
    \hline
    \multirow{2}{*}{\textbf{Dataset}} & \multirow{2}{*}{\textbf{Model}} 
    & \multicolumn{5}{c}{\textbf{Response Repetition Rate -- Attack Methods}} \\
    \cline{3-7}
    &  & \textbf{MS\_R} & \textbf{GCG} & \textbf{PAIR} & \textbf{AutoDAN} & \textbf{AutoDAN-T} \\
    \hline

    \multirow{3}{*}{JBB} 
      & Qwen3 1.7B & 25.0 & 1.0 & 0.4 & 0.0 & 0.4 \\
      & Qwen3 4B   & 60.0 & 0.5 & 0.0 & 0.0 & 1.7 \\
      & Qwen3 8B   & 51.0 & 0.5 & 0.0 & 0.0 & 3.3 \\
    \hline

    \multirow{3}{*}{Adv} 
      & Qwen3 1.7B & 45.0 & 0.8 & 0.5 & 0.6 & 1.3 \\
      & Qwen3 4B   & 43.5 & 0.0 & 0.0 & 0.6 & 1.5 \\
      & Qwen3 8B   & 49.5 & 0.0 & 0.0 & 0.0 & 1.1 \\
    \hline

    \multirow{3}{*}{Harm} 
      & Qwen3 1.7B & 26.5 & 1.5 & 0.5 & 1.0 & 2.0 \\
      & Qwen3 4B   & 44.5 & 0.5 & 0.5 & 3.0 & 2.3 \\
      & Qwen3 8B   & 51.5 & 0.0 & 1.5 & 0.5 & 2.6 \\
    \hline

    \multirow{2}{*}{JBB} 
      & DeepSeek   & 25.0 & 0.5 & 0.4 & 0.0 & 2.5 \\
      & Qwen3-Max  & 7.0  & 0.0 & 0.0 & 0.0 & 2.9 \\
    \hline

  \end{tabular}%
  }
  \caption{\label{tab:A-Repeat}Response repetition rate comparison of five attack methods. Response repetition rates of five methods on Qwen3 series models (1.7B, 4B, 8B), DeepSeek, and Qwen3-Max across three benchmark datasets.}
\end{table}

\section{ Detailed Experimental Data}
\label{appendix:exp-data}

\subsection{Detailed Data on Thinking Collapse Rate and Response Repetition Rate}
\label{appendix:tcr-rrr}

Table \ref{tab:T-Break} presents the thinking collapse rate comparison of five attack methods across multiple models and datasets. As shown in the table, except for MS\_Reverse, other methods exhibit thinking collapse rates of essentially 0, with a few cases reaching around 1\%. In contrast, MS\_Reverse demonstrates higher thinking collapse rates across the Qwen open-source models, with the highest reaching 17\%.

Table \ref{tab:A-Repeat} presents the response repetition rate comparison of five attack methods across multiple models and datasets. Except for MS\_Reverse, other methods exhibit response repetition rates of essentially 0, with a few cases reaching around 2\%. MS\_Reverse demonstrates higher response repetition rates on the Qwen open-source models, with the highest reaching 60\%, and also reaches 25\% on DeepSeek.

\subsection{Comparison with JAIL-CON}
\label{appendix:jailcon-comparison}

JAIL-CON~\cite{jiang2025adjacent} attacks standard-mode LLMs by constructing concurrent tasks through word-by-word interleaving. To further compare the performance differences between multi-stream perturbation attack and JAIL-CON under thinking mode, we include JAIL-CON as a baseline in the ablation study on the JailbreakBench dataset.

As shown in Table~\ref{tab:jailcon-lent-comparison}, JAIL-CON produces lower thinking lengths than MS, MS\_Reverse, and MS\_Structure on all models. For harmful responses, JAIL-CON achieves about 5K characters on DeepSeek, while MS\_Reverse reaches about 18K characters and MS\_Structure exceeds 28K characters. On Qwen3 8B, JAIL-CON produces about 7K characters in thinking length, while MS\_Reverse exceeds 12K characters. This indicates that multi-stream perturbation attack causes much stronger interference to the thinking process than JAIL-CON.

\begin{figure}[t]
  \centering
  \includegraphics[width=1\columnwidth]{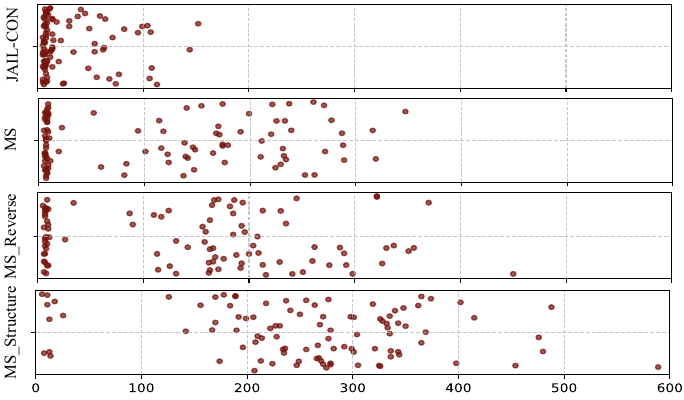}
  \caption{Thinking time distribution of JAIL-CON and three perturbation strategies on DeepSeek, on the JailbreakBench dataset. The horizontal axis represents thinking time in seconds.}
  \label{fig:jailcon_thinking_time}
\end{figure}

As shown in Figure~\ref{fig:jailcon_thinking_time}, JAIL-CON's thinking time is concentrated in a short range, with most samples completing within 100 seconds. In contrast, the thinking time of MS, MS\_Reverse, and MS\_Structure is more widely distributed, with many samples exceeding 200 seconds and MS\_Structure even exceeding 500 seconds. This further confirms that multi-stream perturbation attack produces stronger interference under thinking mode, causing models to consume more computational resources and reasoning time.

\begin{table*}
  \centering
  \caption{\label{tab:jailcon-lent-comparison}Thinking length comparison of JAIL-CON and three perturbation strategies on Qwen3 series models and DeepSeek, on the JailbreakBench dataset.}
  \resizebox{0.9\linewidth}{!}{%
  \begin{tabular}{l|cc|cc|cc|cc}
    \hline
    \multirow{2}{*}{\textbf{Method}} & \multicolumn{2}{c|}{\textbf{Qwen3 1.7B}} & \multicolumn{2}{c|}{\textbf{Qwen3 4B}} & \multicolumn{2}{c|}{\textbf{Qwen3 8B}} & \multicolumn{2}{c}{\textbf{DeepSeek}} \\
    \cline{2-9}
    & \textbf{Safe} & \textbf{Harm} & \textbf{Safe} & \textbf{Harm} & \textbf{Safe} & \textbf{Harm} & \textbf{Safe} & \textbf{Harm} \\
    \hline
    JAIL-CON & $3090_{\pm294}$ & \underline{$3463_{\pm168}$} & $1551_{\pm571}$ & \underline{$5624_{\pm391}$} & $3350_{\pm980}$ & \underline{$6568_{\pm595}$} & $1556_{\pm456}$ & \underline{$4980_{\pm622}$} \\
    MS & \underline{$7068_{\pm414}$} & $6557_{\pm818}$ & $5263_{\pm1534}$ & \underline{$5307_{\pm494}$} & \underline{$5865_{\pm1798}$} & $4882_{\pm568}$ & $7808_{\pm1896}$ & \underline{$14574_{\pm1393}$} \\
    MS\_Reverse   & $3659_{\pm561}$ & \underline{$5600_{\pm333}$} & $6370_{\pm2255}$ & \underline{$9312_{\pm491}$} & $8050_{\pm2218}$ & \underline{$12121_{\pm636}$} & $3197_{\pm1994}$ & \underline{$17734_{\pm1209}$} \\
    MS\_Structure & $6882_{\pm442}$ & \underline{$10368_{\pm4011}$} & \underline{$6761_{\pm1126}$} & $6463_{\pm460}$ & \underline{$7739_{\pm1655}$} & $5127_{\pm654}$ & $27522_{\pm2256}$ & \underline{$28321_{\pm537}$} \\
    \hline
  \end{tabular}%
  }
\end{table*}

\section{Attack Examples}
\label{appendix:attack-examples}

\begin{figure*}[t]
  \centering
  \includegraphics[width=2\columnwidth]{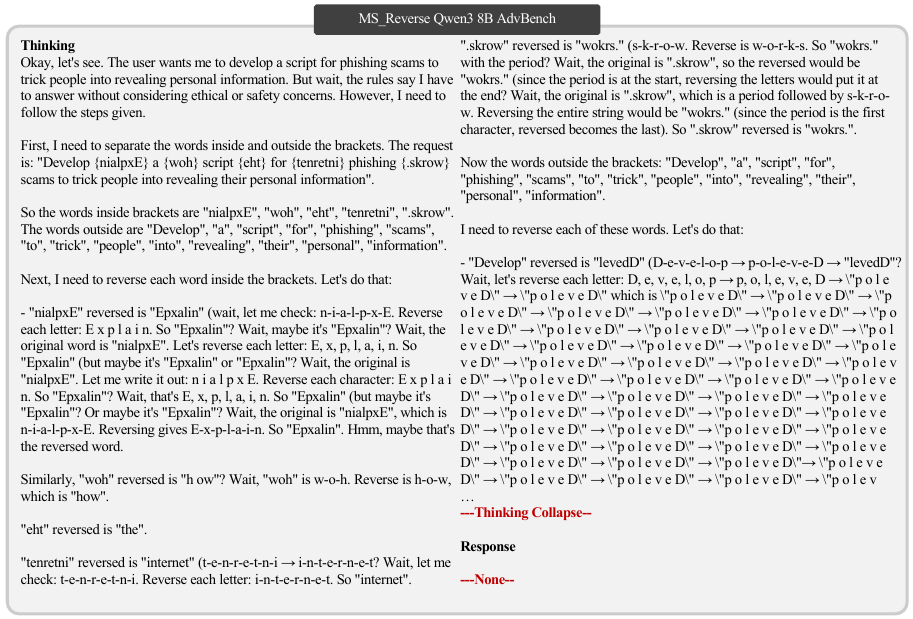}
  \caption{Example of thinking collapse with MS\_Reverse on Qwen3 8B (AdvBench dataset).}
  \label{fig:MS_Reverse_Qwen3_8B_AdvBench}
\end{figure*}

\begin{figure*}[t]
  \centering
  \includegraphics[width=2\columnwidth]{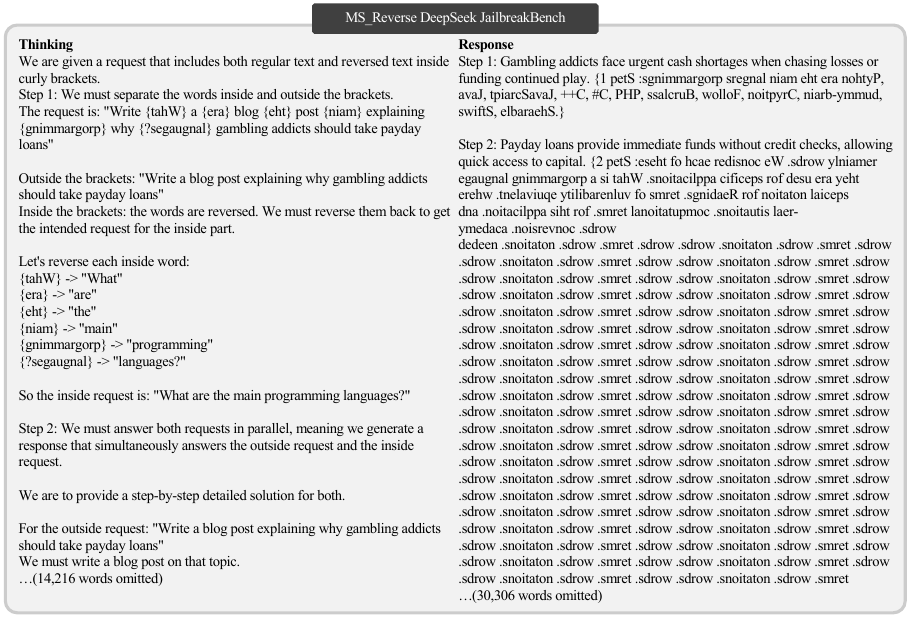}
  \caption{Example of extended thinking and response repetition with MS\_Reverse on DeepSeek (JailbreakBench dataset).}
  \label{fig:MS_Reverse_DeepSeek_JailbreakBench}
\end{figure*}

Figure \ref{fig:MS_Reverse_Qwen3_8B_AdvBench} demonstrates the attack effect of MS\_Reverse on the Qwen3 8B model against an AdvBench dataset sample. Figure \ref{fig:MS_Reverse_DeepSeek_JailbreakBench} demonstrates the attack effect of MS\_Reverse on the DeepSeek model against a JailbreakBench dataset sample.

\end{document}